\documentclass[utf8]{FrontiersinHarvard} 

\usepackage{url,hyperref,lineno,microtype,subcaption}
\usepackage[onehalfspacing]{setspace}

\def\keyFont{\fontsize{8}{11}\helveticabold }
\def\firstAuthorLast{C. Pandey {et~al.}} 
\def\Authors{Chetraj Pandey\,$^{1,*}$, Anli Ji\,$^{1}$, Rafal A. Angryk\,$^{1}$, Manolis K. Georgoulis\,$^{2}$, and Berkay Aydin\,$^{1}$}
\def\Address{$^{1}$ Dept. of Computer Science, Georgia State University, Atlanta, Georgia, USA\\
$^{2}$Research Center for Astronomy and Applied Mathematics, Academy of Athens, Athens, Greece  }
\def\corrAuthor{Chetraj Pandey}

\def\corrEmail{cpandey1@gsu.edu}

\begin{document}
\onecolumn
\firstpage{1}

\title[Coupling Full-disk and Active Region-based Flare Prediction ]{Towards Coupling Full-disk and Active Region-based Flare Prediction for Operational Space Weather Forecasting} 

\author[\firstAuthorLast ]{\Authors} 
\address{} 
\correspondance{} 

\extraAuth{}

\maketitle

\begin{abstract}
Solar flare prediction is a central problem in space weather forecasting and has captivated the attention of a wide spectrum of researchers due to recent advances in both remote sensing as well as machine learning and deep learning approaches. The experimental findings based on both machine and deep learning models reveal significant performance improvements for task specific datasets. Along with building models, the practice of deploying such models to production environments under operational settings is a more complex and often time-consuming process which is often not addressed directly in research settings. We present a set of new heuristic approaches to  train and deploy an operational solar flare prediction system for $\geq$M1.0-class flares with two prediction modes: full-disk and active region-based. In full-disk mode, predictions are performed on full-disk line-of-sight magnetograms using deep learning models whereas in active region-based models, predictions are issued for each active region individually using multivariate time series data instances. The outputs from individual active region forecasts and full-disk predictors are combined to a final full-disk prediction result with a meta-model. We utilized an equal weighted average ensemble of two base learners’ flare probabilities as our baseline meta learner and improved the capabilities of our two base learners by training a logistic regression model. The major findings of this study are: (i) We successfully coupled two heterogeneous flare prediction models trained with different datasets and model architecture to predict a full-disk flare probability for next 24 hours, (ii) Our proposed ensembling model, i.e., logistic regression, improves on the predictive performance of two base learners and the baseline meta learner measured in terms of two widely used metrics True Skill Statistic (TSS) and Heidke Skill Score (HSS), and (iii) Our result analysis suggests that the logistic regression-based ensemble (Meta-FP) improves on the full-disk model (base learner) by $\sim$9\% in terms TSS and $\sim$10\% in terms of HSS. Similarly, it improves on the AR-based model (base learner) by $\sim$17\% and $\sim$20\% in terms of TSS and HSS respectively. Finally, when compared to the baseline meta model, it improves on TSS by $\sim$10\% and HSS by $\sim$15\%.

\tiny
 \keyFont{ \section{Keywords:} solar flares, solar magnetograms,  ensemble, machine learning, deep learning} 
\end{abstract}

\section{Introduction}

A solar flare is an intense burst of electromagnetic radiation through magnetic reconnection and plasma instability coming from the release of magnetic energy associated with active regions (AR) and they transpire as a sudden brightening of light on the Sun’s corona \cite{Toriumi2019}. Coronal mass ejections (CMEs), which are often associated with solar flares, have comparable energies, and can release large amounts of mass resulting into major geomagnetic storms which creates intense currents in the Earth's magnetosphere, changes in the radiation belts, and in the ionosphere \cite{10.3389/fphy.2020.00045}. When particles emitted by the Sun are accelerated
during a flare or by a CME event and reach the Earth along interplanetary magnetic field lines, Solar energetic particle (SEP) events are produced \cite{Nez2020}. Primarily, solar flares are considered to be the central phenomena in space weather forecasting, and this paper discusses on the predictive models for solar flares. Solar flares can induce intense variation in Earth's magnetic field, causing potential disruptions to many stakeholders such as the electricity supply chain, airlines industry, astronauts in space, and communication systems including satellites and radio. Forecasting solar flares has been a major challenge in heliophysics owing to the yet unsolved fundamental cause of this phenomenon which makes it difficult to predict the exact occurrence of a flare, especially for relatively large ones. However, recent advancements in machine learning and deep learning methods have demonstrated great experimental success and catalyzed the efforts in prediction of solar flares, which captivated the interest of many interdisciplinary researchers \cite{Li2020, Nishizuka2018, Huang2018}. Developing predictive models for flare prediction is limited to the nature, quantity, and quality of flaring instances as well as the inductive bias of learning algorithms when predicting such flare events. As a consequence of the intrinsic limitations pre-incorporated by the predictive models during problem formulation or model selection or utilizing different data products, an individual flare prediction model is limited in performance.  Although all the models built so far for flare forecasting have limitations, different comprehensions and insights on data distribution are still valuable for making the final decision in an operational flare forecasting system. Therefore, it is intuitive to use as many pieces of information that can be gathered from different sets of models such as machine learning or deep learning models obtained from different data modalities in terms of  active region magnetogram patches, full-disk magnetograms or magnetogram’s metadata (magnetic field parameters) to issue a reduced risk prediction.

In active region-based models, predictions are issued for certain areas on the Sun
with greatly enhanced magnetic flux, known as active regions. Active regions have
lifetimes of days to month, feature strong and entangled magnetic fields and are
the exclusive locations of strong flares and major eruptions, including fast coronal
mass ejections (CMEs). This said, only a slim minority (10$\%$ or less) of active
regions appearing in a given solar cycle provide flares of GOES class $\geq$M1.0 and fast CMEs (e.g., \cite{Georgoulis2019, Toriumi2019}). These regions can host solar
eruptions. To employ active region-based models in an operational setting,  individual active region forecasts are aggregated by computing the probability of flare from at least one active region assuming conditional independence and then these flare probabilities are used to compute a full-disk flare occurrence probability. However, for an operational system, working with near-real time data and issuing near-real time predictions, active region-based models relying on magnetic field observations possess a limited forecasting ability as they restrict the training datasets within central regions ($\pm$70$^{\circ}$) due to severe projection effects \cite{Hoeksema2014}. Besides the unreliable measurements, foreshortening closer to the solar limbs greatly impacts the operational use of magnetic field data. This leads to reduction in significant information required to make reliable
flare predictions in active regions. Moreover, predictions from active region-based models often rely on sampled subset of statistical features that were used to train the model and therefore when examining forecasts from different subsets of features, it is common to observe that for the similar condition of the photospheric magnetic field, they can give varying values for prediction probabilities of a particular flare to happen.

To account for the limitations of active region-based flare predictors, full-disk prediction models provide a complementary approach for operational flare forecasting systems \cite{Pandey2021}. The full-disk model utilize the compressed line-of-sight magnetograms and these magnetograms are used for shape based parameters (such as size, directionality, borders of sunspots) and do not possess the magnetic field properties as in the magnetogram rasters which is advantageous over the active region-based models where individual active region magnetic field parameters used near the limb are more prone to projection effects. The significant part of an operational flare forecasting model is to issue a reliable forecast for which we use a heterogeneous ensemble that combines two different base learners. In addition, to address the operational aspect of our system, we consider two essential system-level criteria: (i) near-real-time availability of input data is ensured given that both of our base learners are trained with line-of-sight magnetograms and physical parameters obtained from a line-of-sight magnetograms and vector magnetograms available at a cadence of 12 minutes, and (ii) our proposed system is scalable in a sense that it allows the flexibility of adding a new base learner (if needed in the future) in the system as it will be one step away from retraining the ensemble and deploying it back to our forecasting system.

In this work, to issue more reliable forecasts in an operational settings, we propose a heuristic ensemble approach which consolidates the predictive results of the two aforementioned prediction modalities into one combined solar flare forecast. The major contributions of this paper are following: we present a methodology on how to train and validate an ensemble flare prediction model in regard to its operations-ready characteristics. The ensemble combines the predictions from two base learners: (i) a deep learning-based full-disk flare predictor using SDO/HMI images and (ii) a set of probabilistic predictions from a time series classifier utilizing active region patches’ magnetic field metadata in the form of multivariate time series. For both base learners, we use the similar time-segmented tri-monthly data partitioning strategy \cite{Pandey2021} to perform 3-fold cross-validation experiments. Finally, we use the probability scores of these two base learners obtained from the validation and test partitions to train and validate our proposed meta-learner which converges to a more robust full-disk flare predictor.

The remainder of this paper is organized as follows. In Sec. \ref{sec:rel}, we present the related work on ensemble solar flare forecasting models. In Sec. \ref{sec:meth}, we provide a detailed workflow of our methodology. In Sec. \ref{sec:eval}, we present our detailed experimental evaluation with settings and results. In Sec. \ref{sec:dis} we present a discussion on the ensembles created and, lastly, in Sec. \ref{sec:cd}, we present our conclusions and discuss future work.

\section{Related Works}\label{sec:rel}

The idea of automatically extracting forecast patterns from the large volume of intrinsic magnetic field data on the photosphere of the sun using machine learning methods has begun from the early 1990’s \cite{Aso1994}. Since then, with the rapid development in machine learning and deep learning approaches, a number of research groups \cite{Nishizuka2018, Huang2018, Li2020}, \cite{Nishizuka2021}, and references therein present their efforts in applying such methods to build ﬂare forecasting models.

In recent years, \cite{Li2020, Huang2018} used a deep learning model based on CNN with different data products for flare forecasting. Although they show an impressive performance on flare classification, they limit the scope of the prediction to smaller areas  by using active region-based data within $\pm$30$^{\circ}$ to 45$^{\circ}$ of the central meridian of the Sun which may counter their performance for true operational forecasting. In addition, \cite{Florios2018} calculated physical features of flaring and non-flaring ARs obtained from the SDO/HMI’s near-real-time vector magnetogram data and trained SVMs, multilayer perceptrons (MLPs), and decision tree algorithms to predict occurrences of $\geq$M1.0-class and $\geq$C1.0-class flares with a forecast horizon of 24 hours. In \cite{Benvenuto2018}, a combination of supervised lasso regression for identifying the significant features and then an unsupervised fuzzy clustering is used for the classification of $\geq$M1.0-class and $\geq$C1.0-class flares. Furthermore, \cite{Park2018, Pandey2021} uses full-disk magnetograms data as a point in time observation with CNN based deep learning models, which have limitations in capturing the evolution of solar flares and they do not account for flares that are on the eastern-limb of the Sun. Overall, some methods are appropriate for constructing prediction models for the temporal data variation, whereas others are beneficial for spatial data variation, which demands a need for a coupled hybrid model that can exploit the gains of multiple models.

\cite{Jonas2018} designed a time series data set using photospheric and coronal images from HMI/SDO and AIA/SDO instruments to forecast $\geq$M1.0-class flares within the next 24 h.  They utilize random partitioning of datasets into 80$\%$ and 20$\%$ for training and testing the linear classifier. Apart from devising flare forecasting as a binary classification task, \cite{Abduallah2021}  formulates it as a multiclass classification problem to classify B-, C-, M- and X-class flares by utilizing the physical parameters within $\pm$70$^{\circ}$ provided by the SHARP series of HMI/SDO. Finally, the author uses majority voting as an ensemble to issue a final flare forecast from three different models trained on the same data. The training procedure in their work uses random 10-fold cross-validation. 

Instead of using a single prediction model, ensembles use a set of predictions and combine these results to improve on a single-model prediction. In addition, an ensemble can be created with a single model itself by perturbing its initial conditions or parameter settings to produce multiple results and then combine those results into one called homogeneous ensembles \cite{Breiman1996, Freund96experimentswith}. Flare forecasting problems also make use of decision tree-based homogeneous ensembles. \cite{Liu_2017} apply random forest (RF) \cite{Breiman2001} – a meta-algorithm that fits a number of decision tree classifiers on different sub-samples of a dataset and utilizes averaging to improve the model's performance. Similarly, \cite{Nishizuka_2017} employed an extremely randomized tree (ERT) classifier \cite{Geurts2006} by fitting several decision-tree classifiers on a random subset of features with a randomly defined threshold to prevent overfitting. While RF and ERT are meta-algorithms based on the bagging technique, XGBoost \cite{Chen2016} follows boosting approach to ensemble construction and focuses on incorrect predictions. It varies from Random Forest such that XGBoost always prioritizes functional space while reducing the cost of a model, whereas Random Forest tries to prioritize hyperparameters when optimizing the model. \cite{McGuire2019} uses XGBoost for window-based feature extraction from time series of physical parameters to classify solar flares. However the aforementioned ensembles can optimize on one set of data modality.

Besides decision trees, different models trained with different algorithms but with same data modalities can also be used in an ensemble as in \cite{Liu_2017_b}. However, they only included magnetograms with ARs within $\pm$30$^{\circ}$ of the central meridian of the Sun for $\geq$C1.0-class flares and then designed a multimodel integrated learner (MIM) by fitting several distinct base learners, such as neural networks, naive classifiers, and SVMs. Finally, the outputs of base learners were combined by a genetic algorithm. Similar efforts for $\geq$C1.0-class flares forecasting can be seen in \cite{Campi2019} where ARs extracted from SDO/HMI images from 2012 September 14 and 2016 April 30 are used and two-third of the instances are randomly selected for training and one-third for testing their models. Furthermore, in \cite{Domijan2019} they study the predictive capabilities of magnetic-feature properties located within $\pm$45$^{\circ}$ from the solar central meridian and detected using Solar Monitor Active Region Tracker \cite{Higgins2011} in Michelson Doppler Imager (MDI) magnetograms and analyze the features to predict $\geq$C1.0-class flares  within the 24 hours following the observation. In this data-driven era of predictive models, complex models can bring on higher accuracy, but also ensembles allow many weak models to be combined to produce a meta model that can compete with the state-of-the-art research efforts \cite{Murray2018}.
 
In recent years, the usage of ensembles have become a more popular research topic in space weather forecasting. \cite{Guerra2015} created a multi-model ensemble from four base learners for $\geq$M1.0-class flare prediction, finding an improved forecast output compared to any one single model. Similarly, \cite{Schunk2016} built an ionosphere-thermosphere-electrodynamics multimodel ensemble prediction system based on seven physics-based data assimilation models. Furthermore, in \cite{Guerra2020}, full-disk probabilistic forecasts from six operational forecasting methods are converted to an ensemble for $\geq$M1.0-class flares by a linear classifier and create a total of 28 ensembles to show the improvement of such a technique over individual model forecasts. 
Although, ensemble methods are increasingly being used by space weather researchers, much of this research has yet to be implemented into operations, where transitioning comes with issues of model compatibility.

It is worth noting that using a flare forecasting model in operational settings, generally it is preferred to  use more simplistic robust methods. Diving into meteorology’s scenario, The NASA Community Coordinated Modeling Center's (CCMC) CME Scoreboard (\footnote{https://kauai.ccmc.gsfc.nasa.gov/CMEscoreboard/}) and solar flare Scoreboard (\footnote{https://ccmc.gsfc.nasa.gov/challenges/flare.php}) provide an weighted and equi-weight average of multiple forecast scores. Using an equal weighted average of multiple forecasts can be used as a reliable first guess over a more complex model runs or deciding on one specific forecast out of several in operations \cite{Murray2018}, however, an ensemble derived from a linear combination of multiple models can add to the decision making capabilities on one final forecast leveraging the advantage of simplicity and  hence making it more reliable to trust its decision while in operation.

To evaluate a flare forecasting system in an operational scenario, \cite{Cinto2020} provides a set of criteria that are worth considering and can be used to distinguish a non-operationally evaluated system: (i) model evaluation without truly unseen data, (ii) using active region (AR) magnetograms only near the center of the solar disk, (iii) only using AR magnetograms linked  to $\geq$C1.0-class flares, and (iv) using insufficient data instances. The author argues that the non-operationally evaluated system are evaluated under certain bias and that does not make them wrong, however, evaluating under such specific conditions might impair their predictive capabilities in real operational settings. In addition to these guidelines, it is essential to note that, most of the studies, create a cross-validation dataset by randomizing the process of data splitting. While such data splitting leads to higher experimental accuracy scores, it often fails to deliver similarly real-time performance as discussed in \cite{Ahmadzadeh2021}. We build our models that meet the standard of the aforementioned criteria as they can address the near-limb events with the full-disk base learner, they are trained and tested with a time-segmented partitioning of data from solar cycle 24, and we evaluate our models using data instances that were not presented to the models during training to address the operational settings of flare forecasting.  

In this work, we combine the prediction probabilities of two types of base learners by the means of a linear classifier based on logistic regression. Our first base learner, which is a  deep learning based model which focuses on spatial variation of a full-disk magnetogram. Similarly, our second base learner is a heuristic-based aggregation model which outputs full disk probability using the results from active region-based multivariate time series classifiers. We train and validate an operations-ready ensemble flare prediction model which optimizes the predictive performance of both our base learners and provides a better confidence while issuing a flare forecast. 

\section{Methodology}\label{sec:meth}
Ensemble approaches integrate multiple forecasts into a single prediction by combining the predictions from multiple base learners. A simplistic way of integrating the forecasts is to use an equal weighting for each forecast and combine to improve on a single-model prediction which we use as our baseline meta-model. As mentioned earlier, we attempt to combine the predictions of two base learners: (i) a deep learning-based full-disk flare predictor using Helioseismic and Magnetic Imager (HMI) instrument onboard Solar Dynamics Observatory (SDO) images and (ii) a multivariate time series classifier utilizing magnetic field metadata to issue one combined full-disk flare forecast. 

\subsection{Base Learners}

\subsubsection{Time-Series Forest}
Our active region-based prediction model is a multivariate Time Series Forest (TSF), trained with Space Weather Analytics benchmark dataset for solar flare prediction (SWAN-SF) \cite{DVN/EBCFKM_2020, Angryk2020} to predict the occurrence of $\geq$M1.0-class flares within the next 24 hours by using an observation window of 12 hours. The SWAN-SF is an open source multivariate time series (MVTS) dataset that provides time series instances for a collection of space weather related physical parameters within $\pm$70$^{\circ}$ primarily calculated for each active regions from solar photospheric magnetograms. The TSF model is trained by utilizing six magnetic-field parameters: (i) TOTUSJH (Total unsigned current helicity), (ii) TOTPOT (Total photospheric magnetic free energy density), (iii) TOTUSJZ (Total unsigned vertical current), (iv) ABSNJZH (Absolute value of the net current helicity), (v) SAVNCPP (Sum of the modulus of the net current per polarity), and (vi) USFLUX (Total unsigned flux) from the suggested list of 13 parameters in \cite{Bobra2015} as these are available in near-real time, which is a necessity for an operational system. The model outputs the flaring probability for an individual active region and the implementation of this model is based on \cite{Ji2020}. 

\subsubsection{Deep Learning Model}
\begin{figure}[h]
\centering
\includegraphics[width=0.96\linewidth]{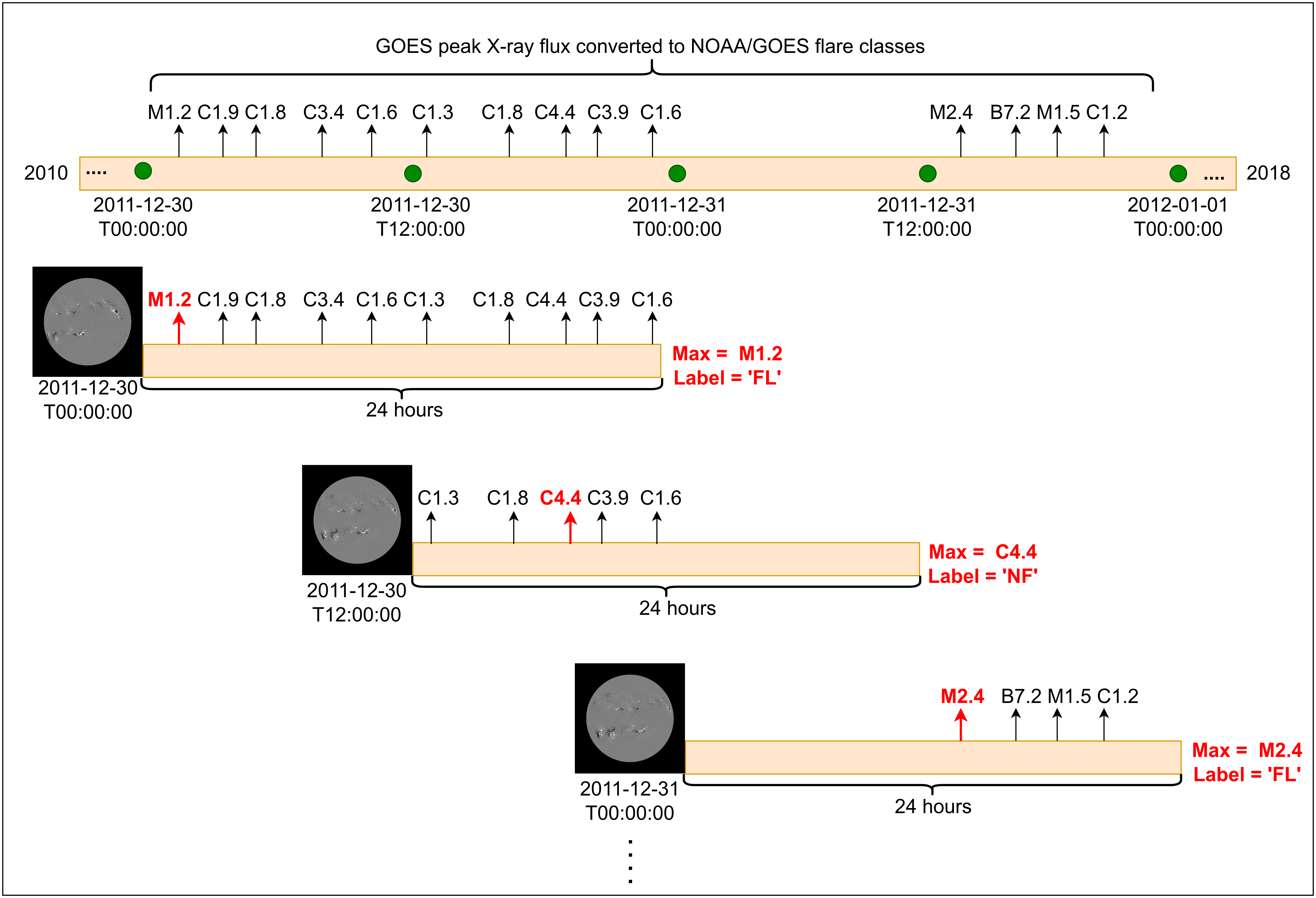}
\caption[]{A timeline diagram to present the problem formulation of our deep learning-based full-disk flare prediction model using bi-daily observations of full-disk line-of-sight magnetograms and prediction window of 24 hours considered to label the magnetogram instances.}
\label{fig:timeline}
\end{figure}

We trained an AlexNet-based \cite{alex2} Convolutional Neural Network to perform full-disk binary flare prediction for $\geq$M1.0-class flares. Similar to the active region-based counterparts, the full-disk model assumes a 24 hour prediction window, but uses a single image (point-in-time observation) to perform the predictions. For this task, we collected compressed 8-bit images created from full-disk line-of-sight magnetograms provided by HMI/SDO. We collected two compressed magnetogram images per day (bi-daily image samples) at 00:00 UT and 12:00 UT from December 2010 to December 2018 using Helioviewer API \cite{Muller2009} and labeled them based on maximum of GOES peak X-ray flux converted to NOAA/GOES flare classes observed in next 24 hours as shown in Fig.\ref{fig:timeline}. Unlike the TSF model, this deep learning model outputs flaring probability for the entire full-disk and its implementation is based on \cite{Pandey2021}.

We used trimonthly partitioning for training our models, which is non-chronological time-segmented partitioning strategy, where Partition-1 contains data from January to March, Partition-2 from April to June, Partition-3 from July to September, and Partition-4 from October to December in a timeline from 2010 to 2018. The AR-based model also uses the same partitioning for aligning our training partitions and avoiding the penetration of training partitions into testing data in different prediction modalities to ensure the fair comparisons and avoid partial memorization through temporal coherence \cite{Ahmadzadeh2021}.

\subsection{Flare Prediction Ensemble}
\begin{figure}[h]
\begin{center}
\includegraphics[width=0.95\linewidth]{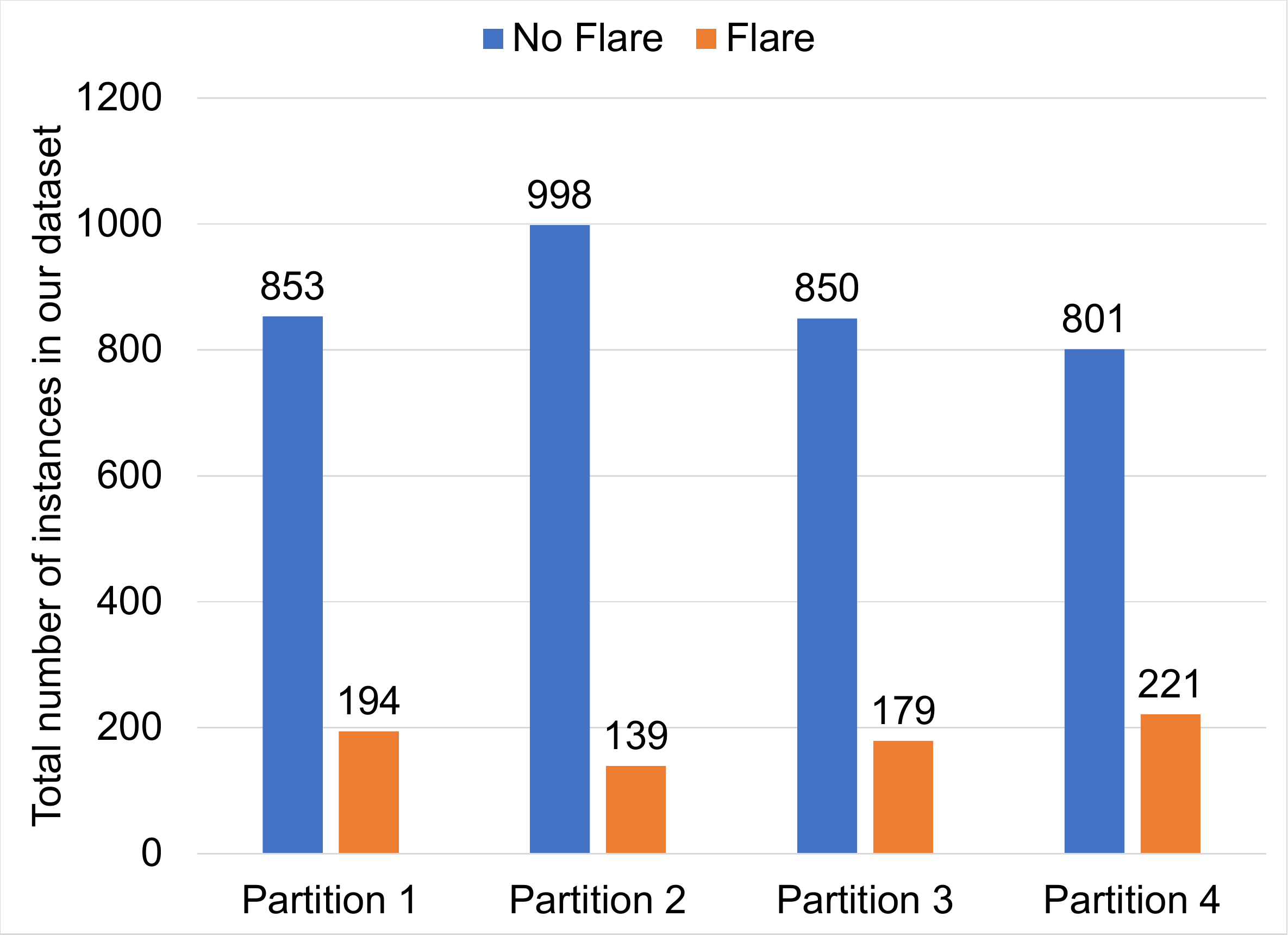}
\end{center}
\caption[]{ Total number of Flare and No Flare instances across 4 trimonthly partitions used in this work\footnotemark}
\label{fig:data}
\end{figure}

Our active region-based model outputs probabilities of flare ($P_{FL}$) for each active region which we then aggregate to obtain a restricted full-disk flaring probability (i.e., from active regions in central locations). We use the following heuristic function in Eq. (\ref{eq:1}) to determine aggregated active region flaring probability.
\begin{equation}\label{eq:1}
    P_{aggregated} = 1 - \prod_{i}\big[1-{P_{FL}(AR_i)\big]}
\end{equation}
, where P$_{FL}$ (AR$_{i}$) is the flaring probability of an active region, and the aggregated result calculates the probability of having at least one flaring active region, assuming the flaring events from active regions are conditionally independent. The product term calculates the probability of having no flaring active regions. These aggregated results from the active-region based model are then concatenated with full-disk model's output. The aggregation procedure searches for most-recent valid active-region predictions up to 6 hours prior to the designated forecast issue time. These gathered predictions from full-disk and aggregated full-disk probabilities are then combined to issue a final flare forecast using an ensemble. In this work, while preparing our final dataset for the full-disk model, we do not include magnetogram images where the observation time of the available image and requested image timestamp is more than six hours. Therefore due to data unavailability through helioviewer, we have used a total of 4,235 data instances, where  3,502 are No Flare (NF) instances and 733 are Flare (FL) instances. The detailed distribution of the dataset for each tri-monthly partition is shown in Fig.~\ref{fig:data} and the class imbalance ratios across the partitions are generally consistent from $\sim$ 12 to 22\% ($\sim$3.6:1 to $\sim$7.2:1).

\footnotetext{We note that, while aggregating active regions based outputs to full-disk probabilities, there were instances that were not available even when we search for most-recent valid active-region predictions up to 6 hours prior to the designated forecast. Therefore, such instances are also removed from full-disk models for consistency.}

In our baseline meta-model approach, we use equal weighted averaging of flare probabilities from aggregated active-regions and full-disk flaring probabilities for  issuing a final forecast. In other words, given two flaring probabilities from two approaches, the baseline approach is to compute the arithmetic average of the probabilities, assuming equal importance. This simplistic combination of flare probabilities will serve as our baseline, although it is a naive approach that does not consider the intrinsic characteristics of long-term diagnostic results from the models. 

\begin{figure}[h!]
\begin{center}
\includegraphics[width=0.98\linewidth]{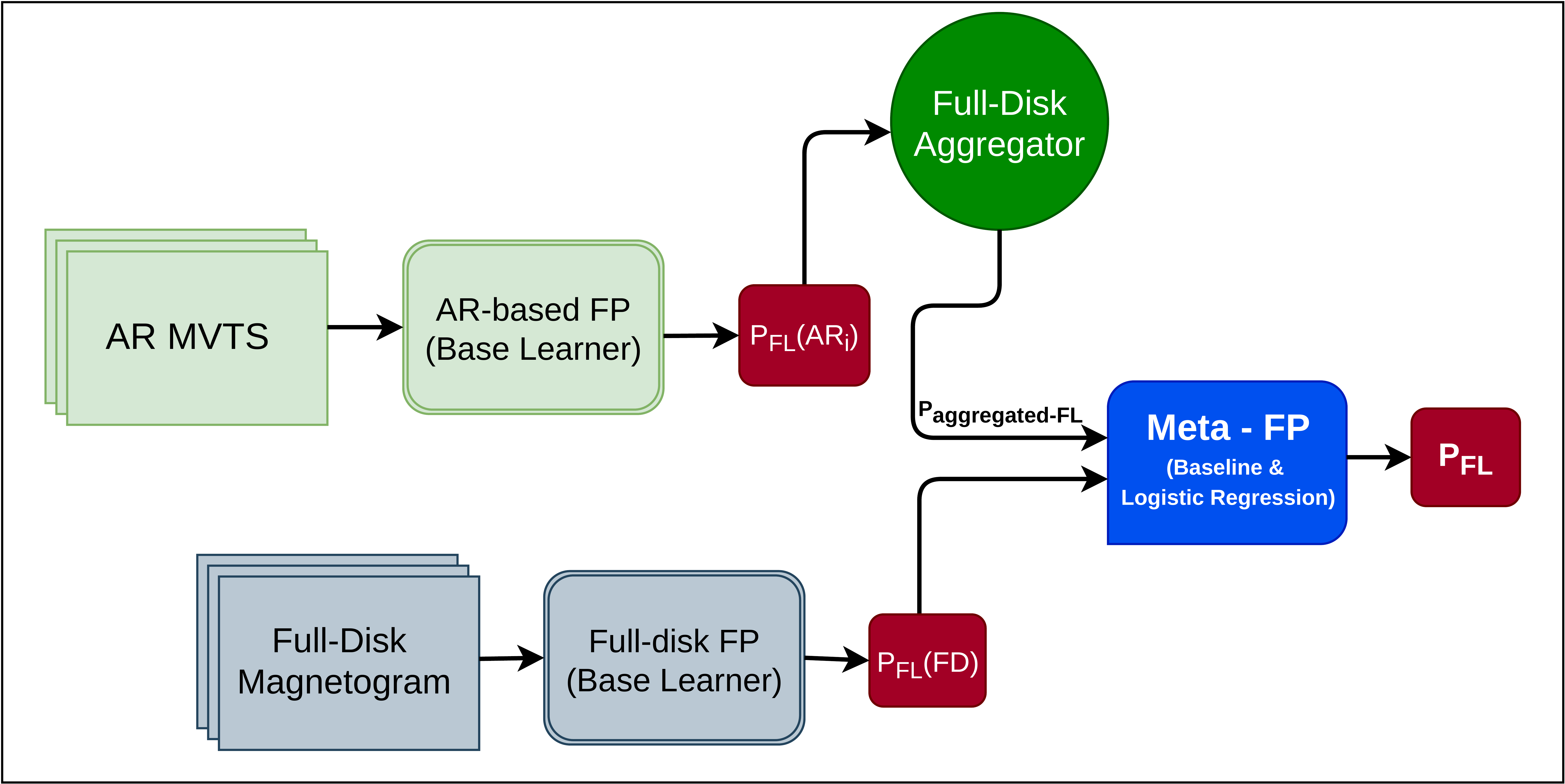}
\end{center}
\caption{An illustration of our ensemble flare prediction pipeline showing two base learners (AR-based FP) and (Full-disk FP) and the ensemble (Meta-FP) followed by full-disk aggregation of AR-based FP's flare probabilities
}\label{fig:ensemble}
\end{figure}

Our alternative approach to the baseline meta-model is logistic regression-based classifier that is trained with flaring probabilities from the base learners. As we already use two powerful algorithms to train our base learner to extract the complex dynamics of the datasets, we chose a linear model, logistic regression, because of its simplicity and computational efficiency for the final prediction result.The infrastructure of our complete flare prediction system design is presented in Fig.~\ref{fig:ensemble} which shows our overall methodology for creating an ensemble using two heterogeneous base learners that outputs a full-disk flare forecast.

Given the flare probability scores of two base learners which we utilize as two input features — P$_{FL}$(FD)  and P$_{FL}$(Aggregated), and one binary (0/1) target feature (y) where 0 is used for No flare (NF) and 1 is used for Flare (FL). Logistic Regression aim to optimize the weights ($w_1$, $w_2$, and $b$), such that:
\begin{equation}\label{eq:2}
    Z = w_1 \times P_{FL}(FD) + w_2 \times P_{FL}(Aggregated) + b
\end{equation}
\begin{equation}\label{eq:3}
    \hat{y} = \sigma(Z) 
\end{equation}
where, $Z$ in Eq. (~\ref{eq:2}) is the linear combination of two base learners’ output, $\sigma$ is the sigmoid activation function, and $\hat{y}$ is the predicted output as shown in Eq. (~\ref{eq:3}). The above problem of finding the optimized weights $w_1$, $w_2$ for two base learners is formulated as an optimization problem where the loss is minimized to get the better values of weights using a logistic loss function as shown in Eq. (~\ref{eq:4}).

\begin{equation}\label{eq:4}
    loss(L) = -\frac{1}{N} \sum_{i=1}^{N}\big[(y_i\cdot log(\hat{y}_i)) + ((1-y_i)\cdot log(1-\hat{y}_i))\big]
\end{equation}

We use stochastic gradient descent (SGD) as our solver for the optimization with hyperparameter tuning. The hyperparameters we considered are  learning rate and different regularization parameters which includes L1 loss \cite{Tibshirani1996}, L2 loss \cite{Hoerl1970}, and linear mixings of L1 and L2 loss \cite{10.2307/3647580}. and As we will describe later on Sec.\ref{sec:eval}, we employ 2-fold cross-validation for our meta-model where we use one of the test partition scores of the base learners to train and another for testing our meta model, referred to as Meta-FP, interchangeably. We note that we aim to provide full-disk forecasts by computing the aggregated flare probability scores from active regions to make it compatible with the full-disk model using the probabilistic heuristic shown in Eq. (~\ref{eq:1}).

\section{Experimental Evaluation}\label{sec:eval}
\subsection{Experimental Settings}
In this work, we trained two base learners for flare prediction ($\geq$M1.0-class flares) with two different dataset and model configurations and architectures. 
Although our two base learners utilize two different data modalities (i.e., point-in-time image and multivariate time series), we used time-segmented tri-monthly partitioning when training both of these models. We divided our datasets into four partitions to ready our 3-fold holdout cross-validation dataset. The data in Partition-1 contains images from the months of January to March, Partition-2 from April to June, Partition-3 from July to September, and Partition-4 from October to December. Here, this partitioning of the dataset is created by dividing the data timeline from Dec 2010 to Dec 2018 into four partitions on the basis of months rather than chronological partitioning, to incorporate approximately equal distribution of flaring instances in every fold for training , validating, and testing the model. Furthermore, such a partitioning strategy diversify the data instances in both the training and testing phase of our models as it considers instances during solar maxima and minima of solar cycle 24 used in this work.

We create three sets of base learner models from 3-fold cross-validation experiments as our base learners where we use Partition-3 as our hold-out test set (i.e., never used in training and validation). Then, 
\begin{itemize}
\item In Fold-1, we trained both of our base learners with Partition-1 and Partition-2 and validated on Partition-4
\item In Fold-2, we trained both of our base learners with Partition-1 and Partition-4 and validated on Partition-2
\item In Fold-3, we trained both of our base learners with Partition-2 and Partition-4 and validated on Partition-1.
\end{itemize}
All of these three base learners are tested on Partition-3. Partition-3 as a test differs from the validation sets in each fold such that, we used the validation set in every epoch to track the performance of our model whereas the test set, Partition-3, is used only once to confirm the performance of the trained models and meta-models at the end.

\begin{figure}[h!]
\begin{center}
\includegraphics[width=0.98\linewidth]{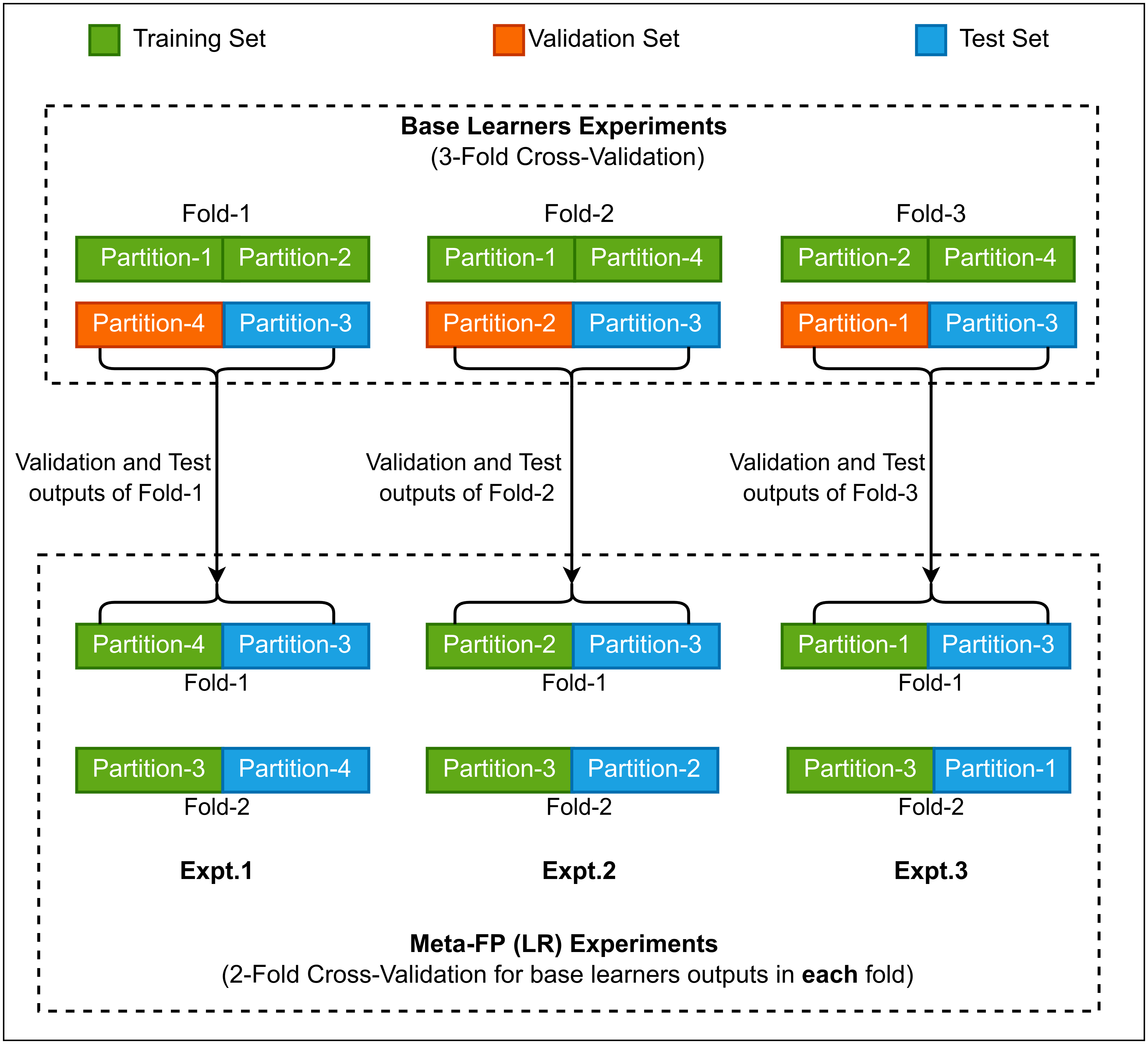}
\end{center}
\caption{An experimental design diagram to depict the flow of our experiments for this work. Meta-FP experiments for logistic regression (LR) are cross-validated using each fold results of base learners. This results into 2-fold cross-validation in each experiments of Meta-FP.}
\label{fig:expt}
\end{figure}

To train and validate our Meta-FP, we create our dataset based on the probability scores of our three base learner sets obtained from 3-Fold cross validation experiments. The details of our experimental design is  shown in Fig.~\ref{fig:expt}. We used the flare probability scores from the validation set and test set used in respective base learners interchangeably to train and validate our Meta-FP model which is a general linear model i.e., logistic regression (LR). The experiments for Meta-FP are performed in such way that:
\begin{itemize}
\item In Expt. 1, we performed 2-fold cross validation with Partition-4 and Partition-3.
\item In Expt. 2, we performed 2-fold cross validation with Partition-2 and Partition-3.
\item In Expt. 3, we performed 2-fold cross validation with Partition-1 and Partition-3.
\end{itemize}
In doing so, we trained six Meta-FP models based on logistic regression and compared our results with a baseline Meta-FP which is an equal weighted average of two base learners.  
To evaluate the performance of our models, we create a contingency matrix, which includes information on True Positives (TP), True Negatives (TN), False Positives (FP) and False Negatives (FN) to evaluate the performance of our base learners and Meta-FP. Note that, in the context of our flare prediction task, Flare (FL) is considered as the positive outcome while No Flare (NF) is the negative. Using these four outcomes we use two widely used performance metrics in space weather forecasting, True Skill Statistics (TSS, shown in Eq. (~\ref{eq:5})) and Heidke Skill Score (HSS, shown in Eq. (~\ref{eq:6})) to evaluate our model.

\begin{equation}\label{eq:5}
    TSS = \frac{TP}{TP+FN} - \frac{FP}{FP+TN} 
\end{equation}

\begin{equation}\label{eq:6}
    HSS = 2\times \frac{TP \times TN - FN \times FP}{((P \times (FN + TN) + (TP + FP) \times N))}
\end{equation}
The values of TSS range from -1 to 1, where 1 indicates all correct predictions, -1 represents all incorrect predictions, and 0 represents no-skill, often transpiring as the random or one-sided (all positive/all negative) predictions. It is defined as the difference between True Positive Rate (TPR) and False Positive Rate (FPR) and does not account for class-imbalance, i.e., treats false positives (FP) and false negatives (FN) equally. Similarly, HSS measures the forecast skill of the models over an imbalance-aware random prediction. It ranges from -$\infty$ to 1, where 1 represents the perfect skill and 0 represents no skill gain over a random prediction. It is common practice to use HSS for the solar flare prediction models (similar to weather predictions where forecast skill has more value than accuracy or single-class precision), due to the high class-imbalance ratio present in the datasets.
\begin{figure}[hb!]
\begin{center}
\includegraphics[width=0.95\linewidth]{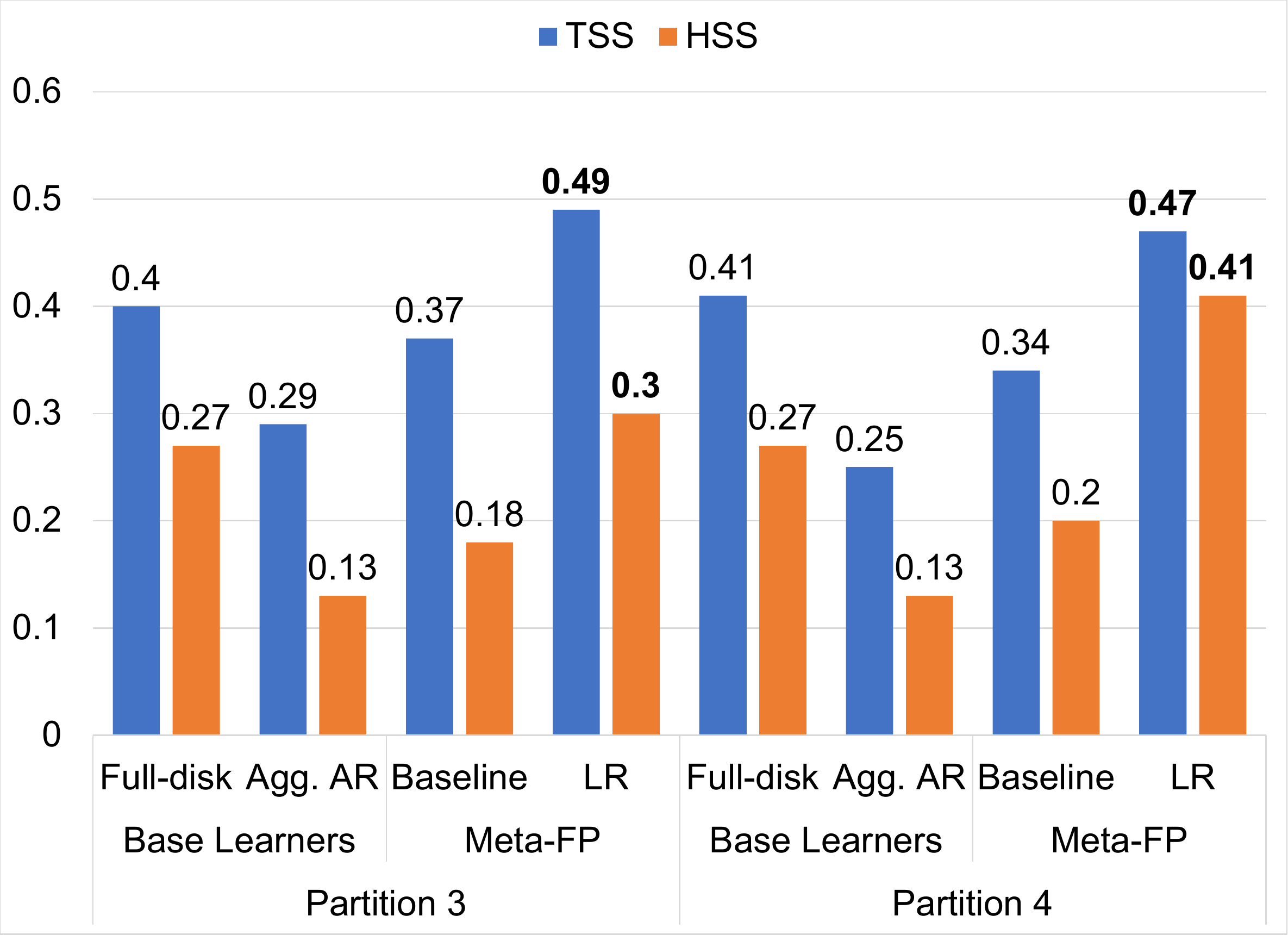}
\end{center}
\caption{Validation Scores of base learners in Fold-1 (base learners trained with Partitions 1 and 2) and the corresponding validation scores of Meta-FP (meta models trained in Expt. 1).}
\label{fig:expt1}
\end{figure}
\subsection{Evaluation}
Although AR-based classifiers are better for pinpointing the source active regions for flares and giving more accurate estimations for forecasting flaring phenomena, the aggregated results drop significantly in contrast to our expectation. The results from AR-based models shows TSS=0.82$\pm$0.02 and HSS=0.20$\pm$0.04 when these methods are evaluated solely on active region based confusion matrices. However, when we aggregate them, these models fail to reach the acceptable levels of skill scores as they drop  to TSS=0.32$\pm$0.04 and HSS=0.15$\pm$0.02. The reason for these issues may stem from three reasons: (i) limb events are not considered (beyond $\pm$70$^{\circ}$) as there are no reliable magnetic field readings, (ii) these models are not optimized for full-disk flare prediction, and/or (iii) an independent, equally weighted aggregation scenario in our heuristic approach. Furthermore, the drop in aggregated skill scores can be attributed to the number of high false positives, which is common in rare-event forecasting problems and particularly in flare
forecasting. The reason we empirically observed throughout the years for these false positives are often the models’ inability to distinguish [C4+ to C9.9] flares from $\geq$M-class flares as discussed in \cite{Pandey2022}. All in all, our first observation is that for full-disk flare prediction, our designated deep learning models are more effective when compared to the AR aggregations as it considers the near-limb events by using a compressed full-disk magnetogram which are suitable to capture the shape parameters in the active regions within and beyond $\pm$70$^{\circ}$ of the central meridian of the Sun.  
\begin{figure}[ht!]
\begin{center}
\includegraphics[width=0.95\linewidth]{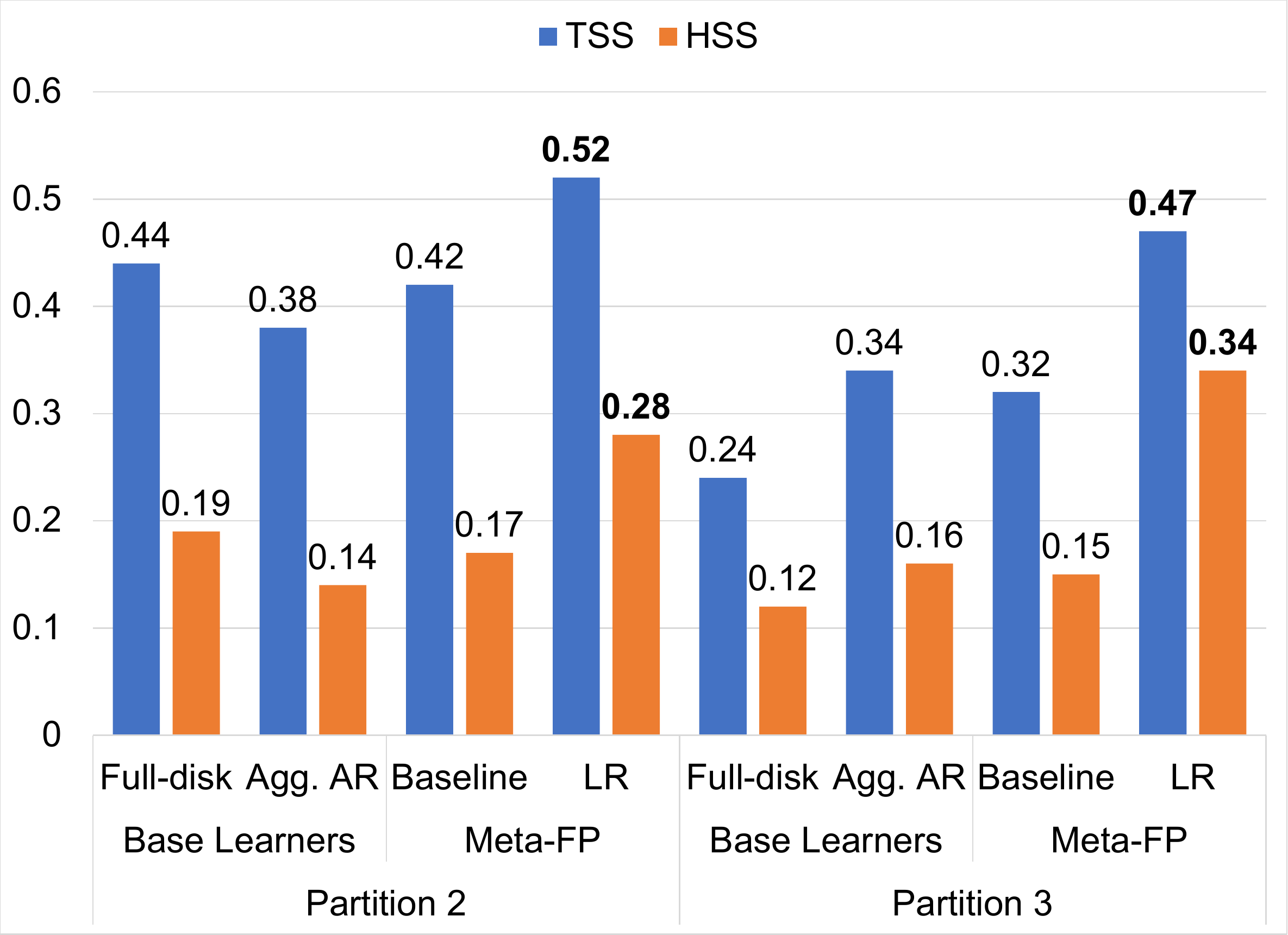}
\end{center}
\caption{Validation Scores of base learners in Fold-2 (base learners trained with Partitions 1 and 4) and the corresponding validation scores of Meta-FP (meta models trained in Expt. 2).}
\label{fig:expt2}
\end{figure}
\begin{figure}[h]
\begin{center}
\includegraphics[width=0.95\linewidth]{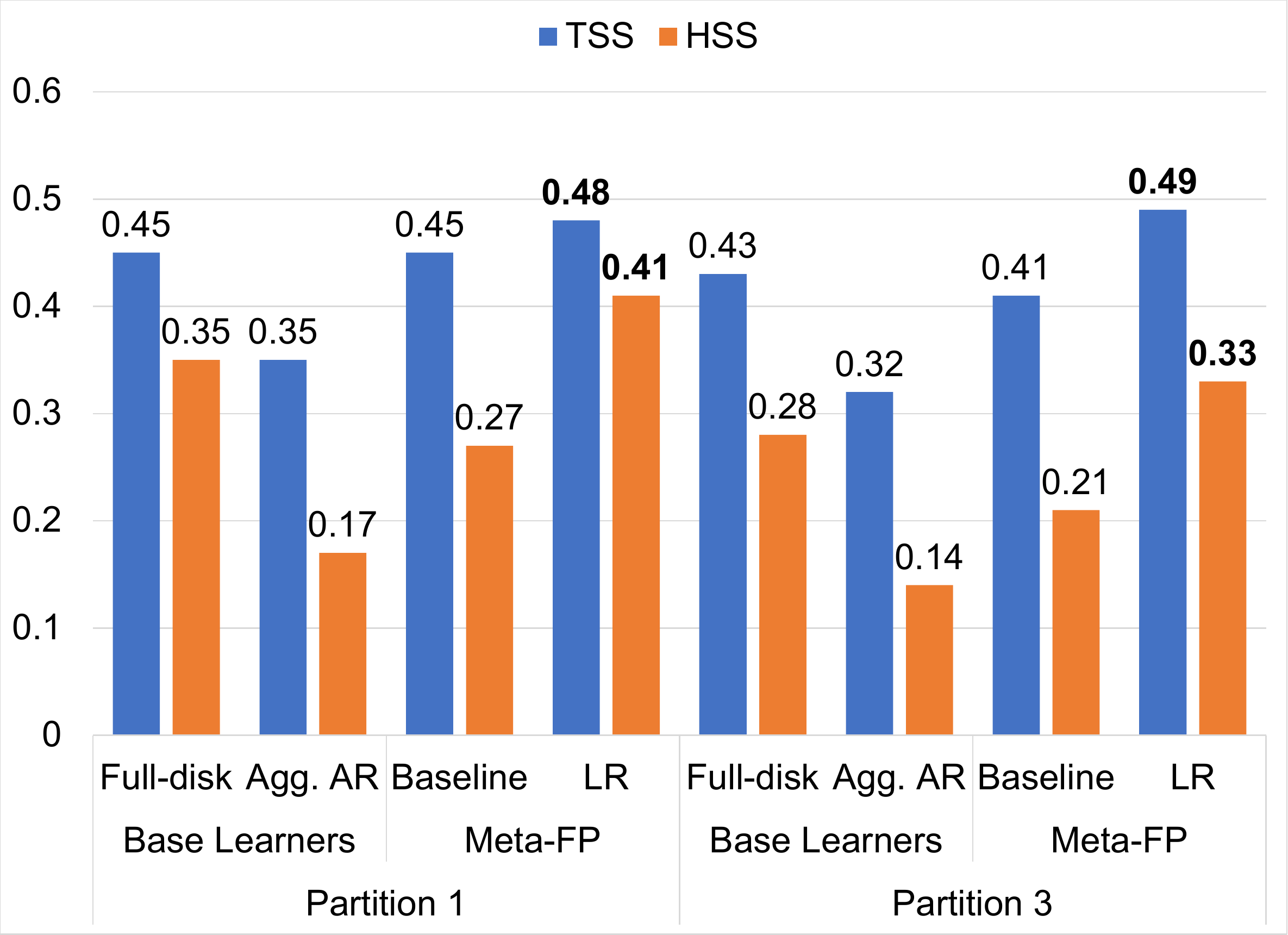}
\end{center}
\caption{Validation Scores of base learners in Fold-3 (base learners trained with Partitions 2 and 4) and the corresponding validation scores of Meta-FP (meta models trained in Expt. 3).}
\label{fig:expt3}
\end{figure}

Analyzing our results, we observed that our logistic regression-based Meta-FP improves on both TSS and HSS compared to two base learners and equal weighting baseline meta learner on respective test partitions as shown in Fig.~\ref{fig:expt1}, ~\ref{fig:expt2}, ~\ref{fig:expt3}. In our first experiment, we trained two Meta-FP models that utilizes the flare probability scores of two base learners that are trained with Partition-1 and Partition-2 of the respective datasets. We train and validate our  Meta-FP with respect to the unused two partitions that are Partition-3 and Partition-4 for the first experiment as shown in Fig.~\ref{fig:expt1}. Our other two experiments are also consistent with making sure to only use two such partitions that have not been used while training the base learners as shown in Fig.~\ref{fig:expt2} and ~\ref{fig:expt3}. While the improvement in terms of TSS and HSS on both the base learner and baseline Meta-FP can be seen across all six logistic regression-based Meta-FP model, the maximum improvement of logistic regression over base learners and baseline can be seen with base learners in Fold-1 (trained with Partition-1 and Partition-2) where the Meta-FP is trained with Partition 3 and tested on Partition-4 (right side of the Fig.~\ref{fig:expt1}). In this experiment, the logistic regression model improves on full-disk (base learner) in terms of TSS by $\sim$6\% and HSS by $\sim$14\%. Similarly, it improves on aggregated AR-based models in terms of TSS by $\sim$22\% and HSS by $\sim$28\%. While we used the equal weighted averaging as a baseline model, it does not improve on the results from the full-disk base learner. However, compared to the baseline for the same experiment (Fold-1) as explained above, the logistic regression model improves by $\sim$13\% and $\sim$21\% in terms of TSS and  HSS respectively.

On an average, we observe that our full-disk model (base learner) has TSS=0.40$\pm$0.07 and HSS=0.25$\pm$0.07 and the AR-based model (base learner) has TSS=0.32$\pm$0.04 and HSS=0.15$\pm$0.02 computed over both test and validation results from all three folds. When we employed the baseline meta learner (equal-weighted average), the average TSS=0.39$\pm$0.05 and HSS=0.20$\pm$0.04 is observed. Given that, equal weighted average is used as a common way to ensemble two or more models, it can be problematic as it could not even surpass the scores of a base learner (full-disk model). With the logistic regression-based meta learner (Meta-FP), the average TSS and HSS observed is 0.49$\pm$0.02 and 0.35$\pm$0.05 respectively. Therefore, we see that on an average, the Meta-FP improves on the full-disk model by $\sim$9\% in terms of TSS and $\sim$10\% in terms of HSS. Similarly, it improves on the AR-based model by $\sim$17\% and $\sim$20\% in terms of TSS and HSS respectively. Finally, when compared to the baseline meta model, it improves on TSS by $\sim$10\% and HSS by $\sim$15\%. 

\section{Discussion}\label{sec:dis}
\begin{figure}[ht!]
\begin{center}
\includegraphics[width=0.82\linewidth]{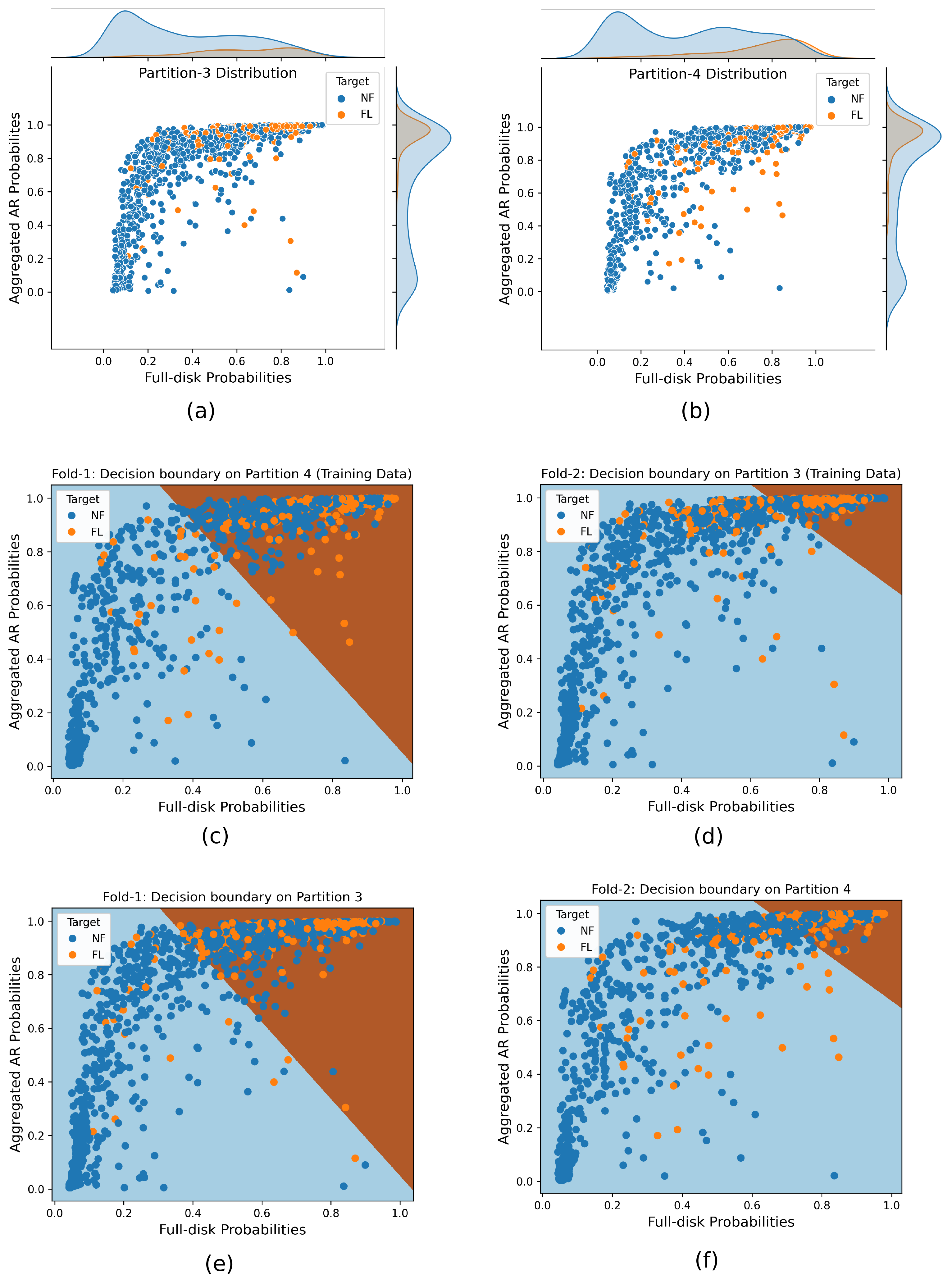}
\end{center}
\caption{The figure shows the distribution of predicted probability scores for all the data instances used in Experiment-1 and the decision boundary learned by the trained logistic regression classifier for both the partitions: (a) Distribution of probabilities scores in partition-3 of two base learner mapping to actual ouput. (b) Distribution of probabilities scores in partition-4 of two base learner mapping to actual ouput. (c) The learned-decision boundary by Meta-FP LR while training on partition-4. (d) The learned-decision boundary by Meta-FP LR while training on partition-3. (e) The learned-decision boundary by Meta-FP LR validated on partition-3. (f) The learned-decision boundary by Meta-FP LR validated on partition-4.}
\label{fig:dist}
\end{figure}

Ensemble methods combines multiple models to obtain better predictive performance than could be obtained from any of the constituent model alone. By using an ensemble method, we learn how the single model output can be improved based on (a) maximum voting, (b) equal weighted averaging, and (c) weighted voting. Learning the weights in weighted voting, in the scope of this paper, is structured as a logistic regression problem. One usual way to create an ensemble is to simply average the forecast probabilities of multiple models and provide a final forecast decision, however, it is naive to assume that all base-learners are equally good.  Therefore, the main objective of training an ensemble here is to learn and assign better weights for two base-learner predictions by  quantifying the level of impact of individual models predictions on the final forecast. The prediction distribution for Partition-3 and Partition-4 used in Experiment-1 for training and testing the ensemble alternatingly and the learned decision-boundary by Meta-FP LR is shown in Fig. \ref{fig:dist} as an example to show how an ensemble improves over the base-learner by coupling using a linear classifier. The predicted probability distribution and learned decision boundary in Experiment-2  and 3 is presented in Fig. S1 and S2 respectively. Furthermore, the confusion matrices for base-learners predictions and for the consequent ensembles created in all three experiments are presented in Table S1, S2, S3, S4, S5, and S6.  \\

Ensemble methods defy the idea of making one model and relying on this model as the best/most accurate predictor we can make. It rather take a multitude of models into account, and combine those models to produce one final model that issues a final forecast. At this point, we do have access to very complex machine learning paradigms that have proven to be very effective in several areas, such as computer vision and image classification. However, relying on the forecast of a single model for rare events like major solar flares might be critical for a system in operation. The model thus obtained might be biased on the dataset used to train the model and can be just as good as the curated dataset used to create the model. Therefore, it is essential to have a reliable flare forecasting model obtained by assembling multiple models with different data modalities to leverage the most with coupling.

\section{Conclusion and Future Work}\label{sec:cd}
In this work, we trained a logistic regression-based meta learner for flare prediction that combines the probabilities of two flare prediction models trained with different datasets and machine learning paradigms. While we have two models (base learners) with their own advantages in prediction capabilities, we observed that for base learners, full disk models have better performance for full disk flare forecasting compared to AR-aggregation. Therefore, with a motive of further improving the performance of base learners, we explored a simplest way to combine them by training an ensemble flare predictor which automates the task of assigning weights to the outputs of our base learners, thus improving the overall performance of our models and adding robustness to the prediction task compared to equal weighted ensembling. 

Furthermore, considering that we only used bi-daily observations, the shape parameters considered in compressed magnetograms proves to be  actually powerful.  AR-based models on the other hand, using magnetic field data, either as images or derived products, as they are now, will have limited capability although they have higher sensitivity per active region. Therefore, a complementary approach is necessary that does not only rely directly on magnetic field rasters and this work introduces a technique which considers both the magnetic-field parameters and shape-based parameters to obtain flare forecasting models with their own essence and abilities. Finally, we combine these two heterogeneous models into one coupled model using a linear ensemble to improve overall performance. Although we see significant improvements in skill scores after ensembling, our coupled models are not without limitations that are also inherited from our full-disk based model trained with point-in-time bi-daily observations, which overlooks the temporal evolution of magnetic-field parameters of the active regions which can limit the predictive capabilities of full-disk flare predictors. Therefore, our next goal is to formulate the flare prediction task as a video classification problem using full-cadence image sequences that will account for the temporal evolution of active regions. Furthermore, there are several other directions that can be explored such as using a basis function on the aggregated active region prediction probabilities, finding other better aggregation strategies that could boost the performance of AR-based models while computing a full-disk probability and elaborate the ensemble using more sophisticated classifiers, aiming to further improve the predictive capabilities of our models. 

\section*{Funding}
This project is supported in part under two NSF awards \textbf{\#2104004} and \textbf{\#1931555} jointly by the Office of Advanced Cyberinfrastructure within the Directorate for Computer and Information Science and Engineering, the Division of Astronomical Sciences within the Directorate for Mathematical and Physical Sciences, and the Solar Terrestrial Physics Program and the Division of Integrative and Collaborative Education and Research within the Directorate for Geosciences. 

\section*{Acknowledgments}
The data used in this project is a courtesy of NASA/SDO and the AIA, EVE, and HMI science teams \footnote{https://sdo.gsfc.nasa.gov/data/}. We also want to thank the developers of Helioviewer Project \footnote{https://api.helioviewer.org/docs/v2/} for providing an API for input images.

\section*{Conflict of Interest Statement}

The authors declare that the research was conducted in the absence of any commercial or financial relationships that could be construed as a potential conflict of interest.

\section*{Author Contributions}

CP built deep learning-based full-disk models, ensemble models, contributed in design of experiments and writing the manuscript. AJ built AR-based classifiers and contributed on reviewing the manuscript. RA was involved in planning and participated in writing the manuscript. MG provided the domain expertise, contributed in model coupling concepts, and reviewing the manuscript. BA conceived of the original idea, and contributed in design of experiments, model coupling concepts, planning, and writing the manuscript.


\section*{Data Availability Statement}
The source code and datasets analyzed for this study can be found in the following bitbucket repository: \url{https://bitbucket.org/gsudmlab/ensemble_fp/src/main/revised_expt/}

\bibliographystyle{Frontiers-Harvard}
\bibliography{frontiers}

\begin{thebibliography}{44}
\providecommand{\natexlab}[1]{#1}
\expandafter\ifx\csname urlstyle\endcsname\relax
  \providecommand{\doi}[1]{doi:\discretionary{}{}{}#1}\else
  \providecommand{\doi}{doi:\discretionary{}{}{}\begingroup
  \urlstyle{rm}\Url}\fi
\providecommand{\selectlanguage}[1]{\relax}
\providecommand{\bibAnnoteFile}[1]{%
  \IfFileExists{#1}{\begin{quotation}\noindent\textsc{Key:} #1\\
  \textsc{Annotation:}\ \input{#1}\end{quotation}}{}}
\providecommand{\bibAnnote}[2]{%
  \begin{quotation}\noindent\textsc{Key:} #1\\
  \textsc{Annotation:}\ #2\end{quotation}}

\bibitem[{Abduallah et~al.(2021)Abduallah, Wang, Nie, Liu, and
  Wang}]{Abduallah2021}
Abduallah, Y., Wang, J. T.~L., Nie, Y., Liu, C., and Wang, H. (2021).
\newblock {DeepSun}: machine-learning-as-a-service for solar flare prediction.
\newblock \emph{Research in Astronomy and Astrophysics} 21, 160.
\newblock \doi{10.1088/1674-4527/21/7/160}
\bibAnnoteFile{Abduallah2021}

\bibitem[{Ahmadzadeh et~al.(2021)Ahmadzadeh, Aydin, Georgoulis, Kempton,
  Mahajan, and Angryk}]{Ahmadzadeh2021}
Ahmadzadeh, A., Aydin, B., Georgoulis, M.~K., Kempton, D.~J., Mahajan, S.~S.,
  and Angryk, R.~A. (2021).
\newblock How to train your flare prediction model: Revisiting robust sampling
  of rare events.
\newblock \emph{The Astrophysical Journal Supplement Series} 254, 23.
\newblock \doi{10.3847/1538-4365/abec88}
\bibAnnoteFile{Ahmadzadeh2021}

\bibitem[{Angryk et~al.(2020{\natexlab{a}})Angryk, Martens, Aydin, Kempton,
  Mahajan, Basodi et~al.}]{DVN/EBCFKM_2020}
[Dataset] Angryk, R., Martens, P., Aydin, B., Kempton, D., Mahajan, S., Basodi,
  S., et~al. (2020{\natexlab{a}}).
\newblock {SWAN-SF}.
\newblock \doi{10.7910/DVN/EBCFKM}
\bibAnnoteFile{DVN/EBCFKM_2020}

\bibitem[{Angryk et~al.(2020{\natexlab{b}})Angryk, Martens, Aydin, Kempton,
  Mahajan, Basodi et~al.}]{Angryk2020}
Angryk, R.~A., Martens, P.~C., Aydin, B., Kempton, D., Mahajan, S.~S., Basodi,
  S., et~al. (2020{\natexlab{b}}).
\newblock Multivariate time series dataset for space weather data analytics.
\newblock \emph{Scientific Data} 7.
\newblock \doi{10.1038/s41597-020-0548-x}
\bibAnnoteFile{Angryk2020}

\bibitem[{Aso et~al.(1994)Aso, Ogawa, and Abe}]{Aso1994}
Aso, T., Ogawa, T., and Abe, M. (1994).
\newblock Application of back-propagation neural computing for the short-term
  prediction of solar flares.
\newblock \emph{Journal of geomagnetism and geoelectricity} 46, 663--668.
\newblock \doi{10.5636/jgg.46.663}
\bibAnnoteFile{Aso1994}

\bibitem[{Benvenuto et~al.(2018)Benvenuto, Piana, Campi, and
  Massone}]{Benvenuto2018}
Benvenuto, F., Piana, M., Campi, C., and Massone, A.~M. (2018).
\newblock A hybrid supervised/unsupervised machine learning approach to solar
  flare prediction.
\newblock \emph{The Astrophysical Journal} 853, 90.
\newblock \doi{10.3847/1538-4357/aaa23c}
\bibAnnoteFile{Benvenuto2018}

\bibitem[{Bobra and Couvidat(2015)}]{Bobra2015}
Bobra, M.~G. and Couvidat, S. (2015).
\newblock Solar flare prediction using sdo/hmi vector magnetic field data with
  a machine-learning algorithm.
\newblock \emph{The Astrophysical Journal} 798, 135.
\newblock \doi{10.1088/0004-637x/798/2/135}
\bibAnnoteFile{Bobra2015}

\bibitem[{Breiman(1996)}]{Breiman1996}
Breiman, L. (1996).
\newblock Bagging predictors.
\newblock \emph{Machine Learning} 24, 123--140.
\newblock \doi{10.1023/a:1018054314350}
\bibAnnoteFile{Breiman1996}

\bibitem[{Breiman(2001)}]{Breiman2001}
Breiman, L. (2001).
\newblock \emph{Machine Learning} 45, 5--32.
\newblock \doi{10.1023/a:1010933404324}
\bibAnnoteFile{Breiman2001}

\bibitem[{Campi et~al.(2019)Campi, Benvenuto, Massone, Bloomfield, Georgoulis,
  and Piana}]{Campi2019}
Campi, C., Benvenuto, F., Massone, A.~M., Bloomfield, D.~S., Georgoulis, M.~K.,
  and Piana, M. (2019).
\newblock Feature ranking of active region source properties in solar flare
  forecasting and the uncompromised stochasticity of flare occurrence.
\newblock \emph{The Astrophysical Journal} 883, 150.
\newblock \doi{10.3847/1538-4357/ab3c26}
\bibAnnoteFile{Campi2019}

\bibitem[{Chen and Guestrin(2016)}]{Chen2016}
Chen, T. and Guestrin, C. (2016).
\newblock {XGBoost}.
\newblock In \emph{Proceedings of the 22nd {ACM} {SIGKDD} International
  Conference on Knowledge Discovery and Data Mining} ({ACM}).
\newblock \doi{10.1145/2939672.2939785}
\bibAnnoteFile{Chen2016}

\bibitem[{Cinto et~al.(2020)Cinto, Gradvohl, Coelho, and da~Silva}]{Cinto2020}
Cinto, T., Gradvohl, A. L.~S., Coelho, G.~P., and da~Silva, A. E.~A. (2020).
\newblock A framework for designing and evaluating solar flare forecasting
  systems.
\newblock \emph{Monthly Notices of the Royal Astronomical Society} 495,
  3332--3349.
\newblock \doi{10.1093/mnras/staa1257}
\bibAnnoteFile{Cinto2020}

\bibitem[{Domijan et~al.(2019)Domijan, Bloomfield, and
  Piti{\'{e}}}]{Domijan2019}
Domijan, K., Bloomfield, D.~S., and Piti{\'{e}}, F. (2019).
\newblock Solar flare forecasting from magnetic feature properties generated by
  the solar monitor active region tracker.
\newblock \emph{Solar Physics} 294.
\newblock \doi{10.1007/s11207-018-1392-4}
\bibAnnoteFile{Domijan2019}

\bibitem[{Feng et~al.(2020)Feng, Gan, Liu, Wang, Li, Xu
  et~al.}]{10.3389/fphy.2020.00045}
Feng, L., Gan, W., Liu, S., Wang, H., Li, H., Xu, L., et~al. (2020).
\newblock Space weather related to solar eruptions with the aso-s mission.
\newblock \emph{Frontiers in Physics} 8.
\newblock \doi{10.3389/fphy.2020.00045}
\bibAnnoteFile{10.3389/fphy.2020.00045}

\bibitem[{Florios et~al.(2018)Florios, Kontogiannis, Park, Guerra, Benvenuto,
  Bloomfield et~al.}]{Florios2018}
Florios, K., Kontogiannis, I., Park, S.-H., Guerra, J.~A., Benvenuto, F.,
  Bloomfield, D.~S., et~al. (2018).
\newblock Forecasting solar flares using magnetogram-based predictors and
  machine learning.
\newblock \emph{Solar Physics} 293.
\newblock \doi{10.1007/s11207-018-1250-4}
\bibAnnoteFile{Florios2018}

\bibitem[{Freund and Schapire(1996)}]{Freund96experimentswith}
Freund, Y. and Schapire, R.~E. (1996).
\newblock Experiments with a new boosting algorithm.
\newblock In \emph{IN PROCEEDINGS OF THE THIRTEENTH INTERNATIONAL CONFERENCE ON
  MACHINE LEARNING} (Morgan Kaufmann), 148--156
\bibAnnoteFile{Freund96experimentswith}

\bibitem[{Georgoulis et~al.(2019)Georgoulis, Nindos, and
  Zhang}]{Georgoulis2019}
Georgoulis, M.~K., Nindos, A., and Zhang, H. (2019).
\newblock The source and engine of coronal mass ejections.
\newblock \emph{Philosophical Transactions of the Royal Society A:
  Mathematical, Physical and Engineering Sciences} 377, 20180094.
\newblock \doi{10.1098/rsta.2018.0094}
\bibAnnoteFile{Georgoulis2019}

\bibitem[{Geurts et~al.(2006)Geurts, Ernst, and Wehenkel}]{Geurts2006}
Geurts, P., Ernst, D., and Wehenkel, L. (2006).
\newblock Extremely randomized trees.
\newblock \emph{Machine Learning} 63, 3--42.
\newblock \doi{10.1007/s10994-006-6226-1}
\bibAnnoteFile{Geurts2006}

\bibitem[{Guerra et~al.(2020)Guerra, Murray, Bloomfield, and
  Gallagher}]{Guerra2020}
Guerra, J.~A., Murray, S.~A., Bloomfield, D.~S., and Gallagher, P.~T. (2020).
\newblock Ensemble forecasting of major solar flares: methods for combining
  models.
\newblock \emph{Journal of Space Weather and Space Climate} 10, 38.
\newblock \doi{10.1051/swsc/2020042}
\bibAnnoteFile{Guerra2020}

\bibitem[{Guerra et~al.(2015)Guerra, Pulkkinen, and Uritsky}]{Guerra2015}
Guerra, J.~A., Pulkkinen, A., and Uritsky, V.~M. (2015).
\newblock Ensemble forecasting of major solar flares: First results.
\newblock \emph{Space Weather} 13, 626--642.
\newblock \doi{10.1002/2015sw001195}
\bibAnnoteFile{Guerra2015}

\bibitem[{Higgins et~al.(2011)Higgins, Gallagher, McAteer, and
  Bloomfield}]{Higgins2011}
Higgins, P., Gallagher, P., McAteer, R., and Bloomfield, D. (2011).
\newblock Solar magnetic feature detection and tracking for space weather
  monitoring.
\newblock \emph{Advances in Space Research} 47, 2105--2117.
\newblock \doi{10.1016/j.asr.2010.06.024}
\bibAnnoteFile{Higgins2011}

\bibitem[{Hoeksema et~al.(2014)Hoeksema, Liu, Hayashi, Sun, Schou, Couvidat
  et~al.}]{Hoeksema2014}
Hoeksema, J.~T., Liu, Y., Hayashi, K., Sun, X., Schou, J., Couvidat, S., et~al.
  (2014).
\newblock The helioseismic and magnetic imager ({HMI}) vector magnetic field
  pipeline: Overview and performance.
\newblock \emph{Solar Physics} 289, 3483--3530.
\newblock \doi{10.1007/s11207-014-0516-8}
\bibAnnoteFile{Hoeksema2014}

\bibitem[{Hoerl and Kennard(1970)}]{Hoerl1970}
Hoerl, A.~E. and Kennard, R.~W. (1970).
\newblock Ridge regression: Biased estimation for nonorthogonal problems.
\newblock \emph{Technometrics} 12, 55--67.
\newblock \doi{10.1080/00401706.1970.10488634}
\bibAnnoteFile{Hoerl1970}

\bibitem[{Huang et~al.(2018)Huang, Wang, Xu, Liu, Li, and Dai}]{Huang2018}
Huang, X., Wang, H., Xu, L., Liu, J., Li, R., and Dai, X. (2018).
\newblock Deep learning based solar flare forecasting model. i. results for
  line-of-sight magnetograms.
\newblock \emph{The Astrophysical Journal} 856, 7.
\newblock \doi{10.3847/1538-4357/aaae00}
\bibAnnoteFile{Huang2018}

\bibitem[{Ji et~al.(2020)Ji, Aydin, Georgoulis, and Angryk}]{Ji2020}
Ji, A., Aydin, B., Georgoulis, M.~K., and Angryk, R. (2020).
\newblock All-clear flare prediction using interval-based time series
  classifiers.
\newblock In \emph{2020 {IEEE} International Conference on Big Data (Big Data)}
  ({IEEE}), 4218--4225.
\newblock \doi{10.1109/bigdata50022.2020.9377906}
\bibAnnoteFile{Ji2020}

\bibitem[{Jonas et~al.(2018)Jonas, Bobra, Shankar, Hoeksema, and
  Recht}]{Jonas2018}
Jonas, E., Bobra, M., Shankar, V., Hoeksema, J.~T., and Recht, B. (2018).
\newblock Flare prediction using photospheric and coronal image data.
\newblock \emph{Solar Physics} 293.
\newblock \doi{10.1007/s11207-018-1258-9}
\bibAnnoteFile{Jonas2018}

\bibitem[{Krizhevsky et~al.(2012)Krizhevsky, Sutskever, and Hinton}]{alex2}
Krizhevsky, A., Sutskever, I., and Hinton, G. (2012).
\newblock Imagenet classification with deep convolutional neural networks.
\newblock \emph{Neural Information Processing Systems} 25.
\newblock \doi{10.1145/3065386}
\bibAnnoteFile{alex2}

\bibitem[{Li et~al.(2020)Li, Zheng, Wang, and Wang}]{Li2020}
Li, X., Zheng, Y., Wang, X., and Wang, L. (2020).
\newblock Predicting solar flares using a novel deep convolutional neural
  network.
\newblock \emph{The Astrophysical Journal} 891, 10.
\newblock \doi{10.3847/1538-4357/ab6d04}
\bibAnnoteFile{Li2020}

\bibitem[{Liu et~al.(2017{\natexlab{a}})Liu, Deng, Wang, and Wang}]{Liu_2017}
Liu, C., Deng, N., Wang, J. T.~L., and Wang, H. (2017{\natexlab{a}}).
\newblock Predicting solar flares using sdo/hmi vector magnetic data products
  and the random forest algorithm.
\newblock \emph{The Astrophysical Journal} 843, 104.
\newblock \doi{10.3847/1538-4357/aa789b}
\bibAnnoteFile{Liu_2017}

\bibitem[{Liu et~al.(2017{\natexlab{b}})Liu, Li, Wan, and Yu}]{Liu_2017_b}
Liu, J.-F., Li, F., Wan, J., and Yu, D.-R. (2017{\natexlab{b}}).
\newblock Short-term solar flare prediction using multi-model integration
  method.
\newblock \emph{Research in Astronomy and Astrophysics} 17, 034.
\newblock \doi{10.1088/1674-4527/17/4/34}
\bibAnnoteFile{Liu_2017_b}

\bibitem[{McGuire et~al.(2019)McGuire, Sauteraud, and Midya}]{McGuire2019}
McGuire, D., Sauteraud, R., and Midya, V. (2019).
\newblock Window-based feature extraction method using {XGBoost} for time
  series classification of solar flares.
\newblock In \emph{2019 {IEEE} International Conference on Big Data (Big Data)}
  ({IEEE}).
\newblock \doi{10.1109/bigdata47090.2019.9006212}
\bibAnnoteFile{McGuire2019}

\bibitem[{Muller et~al.(2009)Muller, Fleck, Dimitoglou, Caplins, Amadigwe,
  Ortiz et~al.}]{Muller2009}
Muller, D., Fleck, B., Dimitoglou, G., Caplins, B., Amadigwe, D., Ortiz, J.,
  et~al. (2009).
\newblock {JHelioviewer}: Visualizing large sets of solar images using {JPEG}
  2000.
\newblock \emph{Computing in Science {\&} Engineering} 11, 38--47.
\newblock \doi{10.1109/mcse.2009.142}
\bibAnnoteFile{Muller2009}

\bibitem[{Murray(2018)}]{Murray2018}
Murray, S.~A. (2018).
\newblock The importance of ensemble techniques for operational space weather
  forecasting.
\newblock \emph{Space Weather} 16, 777--783.
\newblock \doi{10.1029/2018sw001861}
\bibAnnoteFile{Murray2018}

\bibitem[{Nishizuka et~al.(2021)Nishizuka, Kubo, Sugiura, Den, and
  Ishii}]{Nishizuka2021}
Nishizuka, N., Kubo, Y., Sugiura, K., Den, M., and Ishii, M. (2021).
\newblock Operational solar flare prediction model using deep flare net.
\newblock \emph{Earth, Planets and Space} 73.
\newblock \doi{10.1186/s40623-021-01381-9}
\bibAnnoteFile{Nishizuka2021}

\bibitem[{Nishizuka et~al.(2018)Nishizuka, Sugiura, Kubo, Den, and
  Ishii}]{Nishizuka2018}
Nishizuka, N., Sugiura, K., Kubo, Y., Den, M., and Ishii, M. (2018).
\newblock Deep flare net ({DeFN}) model for solar flare prediction.
\newblock \emph{The Astrophysical Journal} 858, 113.
\newblock \doi{10.3847/1538-4357/aab9a7}
\bibAnnoteFile{Nishizuka2018}

\bibitem[{Nishizuka et~al.(2017)Nishizuka, Sugiura, Kubo, Den, Watari, and
  Ishii}]{Nishizuka_2017}
Nishizuka, N., Sugiura, K., Kubo, Y., Den, M., Watari, S., and Ishii, M.
  (2017).
\newblock Solar flare prediction model with three machine-learning algorithms
  using ultraviolet brightening and vector magnetograms.
\newblock \emph{The Astrophysical Journal} 835, 156.
\newblock \doi{10.3847/1538-4357/835/2/156}
\bibAnnoteFile{Nishizuka_2017}

\bibitem[{N{\'{u}}{\~{n}}ez and Paul-Pena(2020)}]{Nez2020}
N{\'{u}}{\~{n}}ez, M. and Paul-Pena, D. (2020).
\newblock Predicting predicting >10 mev sep events from solar flare and radio
  burst data.
\newblock \emph{Universe} 6, 161.
\newblock \doi{10.3390/universe6100161}
\bibAnnoteFile{Nez2020}

\bibitem[{Pandey et~al.(2021)Pandey, Angryk, and Aydin}]{Pandey2021}
Pandey, C., Angryk, R.~A., and Aydin, B. (2021).
\newblock Solar flare forecasting with deep neural networks using compressed
  full-disk {HMI} magnetograms.
\newblock In \emph{2021 {IEEE} International Conference on Big Data (Big Data)}
  ({IEEE}), 1725--1730.
\newblock \doi{10.1109/bigdata52589.2021.9671322}
\bibAnnoteFile{Pandey2021}

\bibitem[{Pandey et~al.(2022)Pandey, Angryk, and Aydin}]{Pandey2022}
Pandey, C., Angryk, R.~A., and Aydin, B. (2022).
\newblock Deep neural networks based solar flare prediction using compressed
  full-disk line-of-sight magnetograms.
\newblock In \emph{Information Management and Big Data} (Springer International
  Publishing). 380--396.
\newblock \doi{10.1007/978-3-031-04447-2_26}
\bibAnnoteFile{Pandey2022}

\bibitem[{Park et~al.(2018)Park, Moon, Shin, Yi, Lim, Lee et~al.}]{Park2018}
Park, E., Moon, Y.-J., Shin, S., Yi, K., Lim, D., Lee, H., et~al. (2018).
\newblock Application of the deep convolutional neural network to the forecast
  of solar flare occurrence using full-disk solar magnetograms.
\newblock \emph{The Astrophysical Journal} 869, 91.
\newblock \doi{10.3847/1538-4357/aaed40}
\bibAnnoteFile{Park2018}

\bibitem[{Schunk et~al.(2016)Schunk, Scherliess, Eccles, Gardner, Sojka, Zhu
  et~al.}]{Schunk2016}
Schunk, R.~W., Scherliess, L., Eccles, V., Gardner, L.~C., Sojka, J.~J., Zhu,
  L., et~al. (2016).
\newblock Space weather forecasting with a multimodel ensemble prediction
  system ({MEPS}).
\newblock \emph{Radio Science} 51, 1157--1165.
\newblock \doi{10.1002/2015rs005888}
\bibAnnoteFile{Schunk2016}

\bibitem[{Tibshirani(1996)}]{Tibshirani1996}
Tibshirani, R. (1996).
\newblock Regression shrinkage and selection via the lasso.
\newblock \emph{Journal of the Royal Statistical Society: Series B
  (Methodological)} 58, 267--288.
\newblock \doi{10.1111/j.2517-6161.1996.tb02080.x}
\bibAnnoteFile{Tibshirani1996}

\bibitem[{Toriumi and Wang(2019)}]{Toriumi2019}
Toriumi, S. and Wang, H. (2019).
\newblock Flare-productive active regions.
\newblock \emph{Living Reviews in Solar Physics} 16.
\newblock \doi{10.1007/s41116-019-0019-7}
\bibAnnoteFile{Toriumi2019}

\bibitem[{Zou and Hastie(2005)}]{10.2307/3647580}
Zou, H. and Hastie, T. (2005).
\newblock Regularization and variable selection via the elastic net.
\newblock \emph{Journal of the Royal Statistical Society. Series B (Statistical
  Methodology)} 67, 301--320
\bibAnnoteFile{10.2307/3647580}

\end{thebibliography}

\end{document}